\documentclass{pasj00}
\usepackage{color}
\draft

\begin{document}
\SetRunningHead{N.Kawano, et al.}{Search for Non-thermal 
X-ray Emission from Abell 3376}
\Received{2007/**/**}%{yyyy/mm/dd}
\Accepted{2008/**/**}%{yyyy/mm/dd}

\title{Constraint of Non-thermal X-ray Emission from the On-going Merger Cluster Abell 
3376 with Suzaku}

%%% begin:list of authors
\author{Naomi \textsc{Kawano}$^1$, Yasushi \textsc{Fukazawa}$^1$, Sho
\textsc{Nishino}$^1$, Kazuhiro \textsc{Nakazawa}$^2$, \\
Takao \textsc{Kitaguchi}$^2$, Kazuo \textsc{Makishima}$^{2,5}$, Tadayuki
\textsc{Takahashi}$^3$, \\
Motohide \textsc{Kokubun}$^3$, Naomi \textsc{Ota}$^3$, Takaya
\textsc{Ohashi}$^4$, Naoki \textsc{Isobe}$^5$, \\
J. Patrick \textsc{Henry}$^6$, and Ann \textsc{Hornschemeier}$^7$} 
%\thanks{Example: Present Address is xxxxxxxxxx}}
\affil{$^1$Department of Physical Science, Hiroshima University, 1-3-1 Kagamiyama, \\
Higashi-Hiroshima, Hiroshima 739-8526}
\email{kawano@hirax7.hepl.hiroshima-u.ac.jp, fukazawa@hirax7.hepl.hiroshima-u.ac.jp}

\affil{$^2$Department of Physics, University of Tokyo, 7-3-1 Hongo,
Bunkyo, Tokyo 113-0033}

\affil{$^3$Department of High Energy Astrophysics, Institute of Space and \\
Astronomical Science (ISAS), 
Japan Aerospace Exploration Agency (JAXA), \\ 3-1-1 Yoshinodai, Sagamihara, Kanagawa 229-8510}

\affil{$^4$Department of Physics, Tokyo Metropolitan University, \\ 1-1
Minami-Osawa, Hachioji, Tokyo 192-0397}

\affil{$^5$Cosmic Radiation Laboratory, The Institute of Physical and
Chemical Research (RIKEN), \\ 2-1 Hirosawa, Wako, Saitama 351-0198}

\affil{$^6$Institute for Astronomy, University of Hawai'i, 2680 Woodlawn
Drive, \\ Honolulu, HI 96822, USA}

\affil{$^7$Exploration of the Universe Division, NASA/Goddard Space
Flight Center (GSFC), \\ Greenbelt, MD, 20771, USA}

%\author{B-Firstname \textsc{B-Familyname}}
%\affil{B-Address of Institute}\email{bbbbb@xxx.xxx.xx.xx}
%\and
%\author{C-Firstname {\sc C-Familyname}}
%\affil{C-Address of Institute}\email{ccccc@xxx.xxx.xx.xx}
%%% end:list of authors

%%% Please use the following style in case that sorting by 
%%% affilation is impossible. 
%
% \author{%
%   D-Firstname \textsc{D-Familyname}\altaffilmark{1}
%   E-Firstname \textsc{E-Familyname}\altaffilmark{1,2}
%   and
%   F-Firstname \textsc{F-Familyname}\altaffilmark{2}}
% \altaffiltext{1}{Address of Institute}
% \email{ddddd@xxx.xxx.xx.xx}
% \email{eeeee@xxx.xxx.xx.xx}
% \altaffiltext{2}{Address of Institute}

%% `\KeyWords{}' always has to be placed before `\maketitle'.
%Do NOT move this preamble from here!

\maketitle

\begin{abstract}
Clusters of galaxies are among the best candidates for particle
acceleration sources in the universe, a signature of which is
non-thermal hard X-ray emission from the accelerated relativistic
particles.  We present early results on Suzaku observations of
non-thermal emission from Abell 3376, which is a nearby on-going
merger cluster.  Suzaku observed the cluster twice, focusing on the
cluster center containing the diffuse radio emission to the east, and
cluster peripheral region to the west.  For both observations, we
detect no excess hard X-ray emission above the thermal cluster
emission.  An upper limit on the non-thermal X-ray flux of
$2.1\times10^{-11}$ erg cm$^{-2}$ s$^{-1}$ (15--50 keV) at the
3$\sigma$ level from a $34\times34$ arcmin$^2$ region, derived with
the Hard X-ray Detector (HXD), is similar to that obtained with
the BeppoSAX/PDS.  Using the X-ray Imaging Spectrometer (XIS) data,
the upper limit on the non-thermal emission from the West Relic is
independently constrained to be $<1.1\times10^{-12}$ erg s$^{-1}$
cm$^{-2}$ (4$-$8 keV) at the 3$\sigma$ level from a 122 arcmin$^2$
region.  Assuming Compton scattering between relativistic particles
and the cosmic microwave background (CMB) photons, the intracluster
magnetic field $B$ is limited to be $>$0.03$\mu$G (HXD) and
$>$0.10$\mu$G (XIS).

\end{abstract}

\section{Introduction}

Particle acceleration is one of the most exciting phenomena in
clusters of galaxies.  In fact, Mpc-scale diffuse synchrotron emission
has been reported for many clusters of galaxies
(e.g. \cite{giovannini93}), implying that the relativistic electrons
surely exist.  A natural origin of acceleration in clusters is merger
shocks.  The electrons, accelerated to relativistic speeds by a strong
cluster merger (e.g. \cite{takizawa00}; \cite{fujita01}), scatter off
the cosmic microwave background (CMB) photons via the inverse Compton
process. This causes non-thermal hard X-ray emission which may exceed
the thermal bremsstrahlung emission of the intracluster medium (ICM)
at energies above several tens of keV.  Studies of inverse Compton
emission in addition to the radio synchrotron are important to measure
the energy densities of the magnetic field and electrons without
assuming equipartition.  Since the cooling time of such high energy
electrons is relatively short, $\sim$10$^8$ yr, they vanish rapidly
during cluster evolution.  Another mechanism to produce a nonthermal
electron population is via proton-proton collisions between the
relativistic protons and the ICM protons.  In this case, the
relativistic electrons can be provided for a long time because the
lifetime of relativistic protons is comparable to that of galaxy
cluster ($\sim H_0^{-1}$).  Such protons may also cause a gamma-ray
emission through $\pi^0$ decays.  Contribution of non-thermal pressure
to the thermal one is also an important question, since it could
affect the cluster mass estimation under the assumption of hydrostatic
equilibrium.  Accordingly, hard X-ray and gamma-ray measurements of
non-thermal cluster emission are important issues and so far many
attempts have been performed.  Furthermore, the launch of the GLAST
satellite scheduled in 2008 will improve the GeV gamma-ray survey in
the near future.

Some detections of non-thermal hard X-rays of varying levels of
significance have been reported by the BeppoSAX/PDS, ASCA/GIS, and
RXTE/PCA.  Individual detections have been reported of the Coma
cluster (\cite{fusco-femiano99}), Abell 2256 (\cite{fusco-femiano00}),
Abell 2199 (\cite{kaastra99}), Abell 3667 (\cite{fusco-femiano01}),
and HCG 62 (\cite{fukazawa01}, \cite{nakazawa07}).  From the
systematic analysis of 27 clusters using BeppoSAX PDS data, a possible
excess signal of non-thermal emission was reported for 7 galaxy
clusters with $>$2 $\sigma$ level (\cite{nevalainen04}).  The reported
hard X-ray luminosity of these clusters is in a range of
(0.17$-$42)$\times$10$^{43}$ erg s$^{-1}$, which corresponds to 1$-$20
\% of the thermal luminosity.  The spatial distribution is as
important to understand the non-thermal view of galaxy clusters as is
spectral information. However, the spatial extent of the non-thermal
hard X-ray emission has not been measured yet.
 
Abell 3376 (DC 0559-40) is a nearby ($z=$0.046), rich cluster of
galaxy.  A binary subcluster merger is occurring in this cluster, and a
weak positive hard X-ray signal (2.7 $\sigma$) with the
BeppoSAX/PDS was reported by \citet{nevalainen04}.  Another remarkable
feature of this cluster is a pair of Mpc scale radio relics
(\cite{bagchi05}; \cite{bagchi06}).  The 1.4 GHz radio flux densities
of two relics from the VLA NVSS are 32$\pm$3 mJy and 82$\pm$5 mJy for
the east (ER) and the west relic (WR), respectively.  Radio relics are
considered to be synchrotron emission generated through the
interaction between electrons accelerated by merger shock and magnetic
field in the ICM, implying that the relativistic electrons really
exist in this cluster. The giant ringlike structure of the radio relics
imply a strong shock wave which could accelerate particles up to
$10^{18-19}$ eV \citep{bagchi06}.  The thermal properties of this
cluster have been observed with ROSAT, ASCA, Chandra, and XMM-Newton.
The ICM temperature reported by XMM-newton (\cite{bagchi06}) is
moderately low ($\sim$4 keV), similar to that of the ASCA observation
(\cite{fukazawa04}), and we can therefore search for the non-thermal
emission by avoiding thermal component in the energy band above $10$
keV.  From these properties, Abell 3376 emerged as a good candidate to
search for the non-thermal hard X-ray emission.  In this paper, we
present early results of Suzaku observation of Abell 3376, focusing on
the non-thermal hard X-ray emission.

The Si-PIN diode (PIN) in the Hard X-ray detector (HXD;
\cite{takahashi07}; \cite{kokubun07}) onboard Suzaku has a very low
background in the 10$-$30 keV band; the background count rate
normalized by the effective area is smaller than any past instrument.
In addition, its narrow field of view (34$'\times$34$'$ full width at
half maximum, FWHM), compared with the BeppoSAX/PDS (1.3$^{\circ}$
hexagonal FWHM) and RXTE/PCA (1$^{\circ}$ hexagonal FWHM), yields
several advantages.  First, the confusion from hard point sources is
reduced, as is the contribution of the ICM thermal emission, and
finally the non-thermal emission may be localized better.  Suzaku also
employs the X-ray Imaging Spectrometer (XIS; \cite{koyama07}), which
consists of one back-illuminated (BI) CCD chip (XIS1), and three
front-illuminated (FI) CCD chips (XIS0, 2, 3).  Since the XIS also
achieves the lowest background level of any previous X-ray CCD, it is
useful to constrain the hard X-ray emission.

Throughout this paper, we adopt a Hubble constant of $H_0=$50 $h_{50}$
km s$^{-1}$ Mpc$^{-1}$, which implies that 1$'$ corresponds to 80.3
kpc in the cluster rame. All statistical errors are
given at the 90 \% confidence level unless otherwise indicated.

\section{Observations and Data Reduction}

\begin{figure}[ht]
\begin{center}
%\FigureFile(9.0cm,6.0cm){figure/a3376merge.ps}
%\FigureFile(9.0cm,6.0cm){figure/a3376grid.ps}
\FigureFile(9.0cm,6.0cm){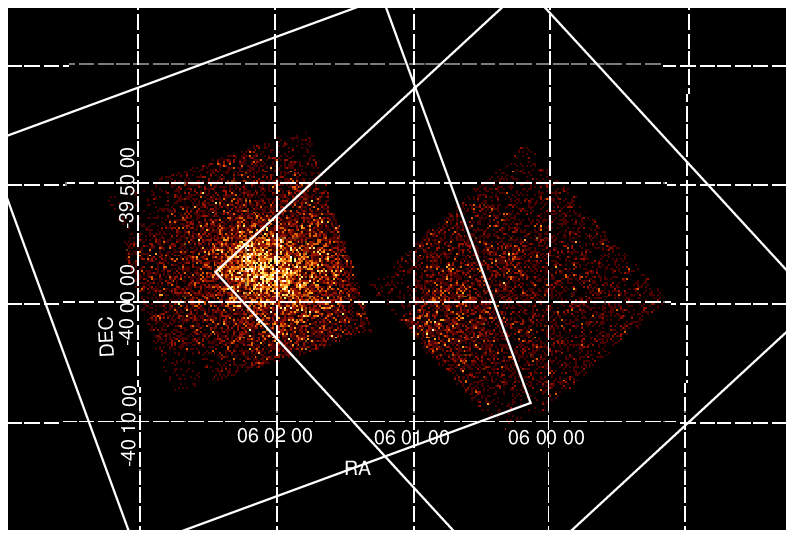}
\end{center}
\caption{Suzaku XIS image of Abell 3376 in the 0.5--8 keV band. All the BI and FI
 data from the two pointings are combined. The corner regions
 illuminated by the calibration source are excluded. The large squares
 represent the HXD-PIN field of view with a FWHM of $34'\times34'$.}
\label{image}
\end{figure}

Two long observations of Abell 3376 were carried out on October 6--10
and November 7--10, 2005, in Suzaku Phase-I period (i.e. initial
performance verification period). The former observation included the
cD galaxy and the ER.  The latter one focused on the WR, which is
$\sim$25' west of the cD galaxy.  The details of each observation are
shown in Table \ref{obslog}.  Both the XIS and HXD were operated in
the normal mode.  Figure \ref{image} shows an X-ray image of Abell
3376 obtained with the XIS in the 0.5$-$8.0 keV band.  The thermal
emission has an X-ray peak around the cD galaxy and is elongated due
to the subcluster merging as observed with previous satellites
(\cite{bagchi05}; \cite{bagchi06}).

We used version 1.2 pipeline processing data (\cite{mitsuda07}), and
the data reduction was performed with {\tt HEAsoft} version 6.2.  For
the HXD data, events are filtered with a cut-off rigidity (COR) of $>$
6 GeV c$^{-1}$, elevation angle of $>$ 5$^{\circ}$ from the earth rim,
and good time interval (GTI) during which the satellite is outside the
South Atlantic Anomaly (SAA).  Moreover, we eliminate the periods
($\sim$500 s) when the PIN count rate is flaring due to in-orbit
particles.  After these procedures, a net PIN exposure time of 90 ks
and 103 ks was obtained for the Center/ER and WR pointing,
respectively.  We applied {\tt ae\_hxd\_pinhxnom\_20060419.rmf} for
the HXD response matrix.  The PIN detector background is estimated by
our own method rather than the standard model, as described in \S3.1.
We do not present the GSO data in this paper, since no significant
signal is detected.

Selection criteria are different for the XIS compared to the HXD only
in the respect that the earth elevation angle of $>$ 20$^{\circ}$ is
applied.  We also examined the 0.5$-$8 keV light curve with 128 s time
bin to remove the flaring periods of the event rate higher than
$\pm$3$\sigma$ above the average.  The resulting exposure times become
120 ks and 127 ks for the Center/ER and WR observations, respectively.
We generated the response matrix files with the {\tt xisrmfgen}
F{\small TOOLS} task, and the auxiliary response files with the {\tt
xissimarfgen} F{\small TOOLS} task (\cite{ishisaki07}).  The latter
takes into account the vignetting effect and corrects the efficiency
degradation in the lower energy band due to contamination on the
optical blocking filters in front of the XIS \citep{koyama07}.  The
ROSAT PSPC image of Abell 3376 was used as the seed image in the {\tt
xissimarfgen}, and also the regions of extracted spectra are taken
into account.  The detector background is estimated by the F{\small
TOOLS} task {\tt xisntebgdgen} (\cite{tawa07}).  We performed spectral
fits in the energy range of 0.3$-$8.0 keV and 0.5$-$8.0 keV for the
XIS BI and the FI, respectively.  The energy range of 1.7$-$1.9 keV is
ignored in the spectral fitting of XIS because the response matrix
around Si-K edge has some residual uncertainties.  All the spectral
analyses were performed with X{\small SPEC} 11.2.0.

\begin{table}[ht]
\caption{Suzaku observation log of Abell 3376}
\label{obslog}
\begin{center}
\small
\begin{tabular}{ccccc}
\hline\hline
Position & Sequence No. & Date & RA, DEC & Exposure time$^*$ \\
\hline
Center/East Relic & 100034010 & 2005 Oct. 6--10 
& 06$^{\rm h}$02$^{\rm m}$15.0$^{\rm s}$, -39$^{\circ}$57$'$00.0$''$ 
& 90 ks / 120 ks \\
West Relic        & 800011010 & 2005 Nov. 7--10 
& 06$^{\rm h}$00$^{\rm m}$00.0$^{\rm s}$, -40$^{\circ}$01$'$58.8$''$ 
& 103 ks / 127 ks \\
\hline
\multicolumn{5}{l}{*: The net exposure time of HXD/XIS after the events are screened.} 
\end{tabular}
\end{center}
\normalsize
\end{table}

\section{Data Analysis and Results}

\subsection{Spectral Analysis}

\subsubsection{HXD Spectra\label{s-hxdsp}}

First, we explore the non-thermal emission with the HXD PIN spectrum.
Five components are here considered; 
ICM thermal emission, cosmic
X-ray background (CXB), non-X-ray detector background (NXB), point
sources, and possible non-thermal emission. 
Among these components, NXB contribution to the total counts is
the highest. Therefore, it is most important 
to predict the NXB as accurately as possible for
detection of faint non-thermal emission. 
The HXD team has officially supplied the NXB model for the PIN, which
reproduces the NXB by sorting the earth-occultation data-base with two
parameters; PINUD count and PINUD build-up count.
The latter is a convolution of PINUD with an exponential function with
an appropriate decay constant, in order to represent the
activation-induced gamma-rays due to the SAA passage.
Since this component is non-negligible for a day or so after
the SAA passage,
the PINUD count history just before the observation is needed
to model the background of that observation.
Unfortunately, the satellite telemetry data was not received for almost 
a day before the observation of A3376 WR (Nov 7--10, 2005), 
consequently the PINUD data are not available for that time period.
Thus, the official NXB background model is not accurate, and 
here we apply an alternative background estimation.
Our estimation method is based on the NXB count rate map sorted by
latitude and longitude on the earth for the satellite position. 
We used earth occultation data as the NXB template.
After the long-term variation is
corrected, the NXB uncertainty finally becomes $\sim$2\% at the 1$\sigma$
level, which is a
similar reproducibility to that of the official NXB model. 
The details of our NXB estimation method and its systematic uncertainty 
are described in Appendix \ref{nxb}.
Further details are given in \citet{kawano06}.
As a result, our NXB is consistent with the official
NXB within their systematic errors even for the WR observation.

We examine the possible non-thermal emission by estimating the
contribution of the other components to the NXB-subtracted spectra as
follows.  First for the ICM flux we used that within 19 arcmin
measured by the ASCA/GIS, 2.5$\times10^{-11}$ erg cm$^{-2}$ s$^{-1}$
(0.5--10 keV) (\cite{fukazawa04}).
Since XIS detected the emission up to 22 arcmin, we extrapolated the
above flux to 22 arcmin
with a $\beta$-model with parameters obtained from the same ASCA/GIS
data, yielding 3.0$\times10^{-11}$ erg cm$^{-2}$ s$^{-1}$ (0.5--10
keV). We used this value to normalize the X-ray surface brightness
from an archival ROSAT/PSPC image, then folded the result through the
PIN collimator transmission.  The result is the total ICM flux that
would be observed by the PIN if it were sensitive in the 0.5--10 keV
band: $2.1\times10^{-11}$ erg cm$^{-2}$ s$^{-1}$ and
$1.5\times10^{-11}$ erg cm$^{-2}$ s$^{-1}$ for the Center/ER and WR
observation, respectively.  We extrapolated the spectrum into the PIN
energy band assuming the spectral parameters in \citet{fukazawa04}, obtaining
$2.5\times10^{-13}$ erg cm$^{-2}$ s$^{-1}$ and $1.8\times10^{-13}$ erg
cm$^{-2}$ s$^{-1}$ (15--50 keV) for the Center/ER and WR observation,
respectively.  In order to check the consistency between PSPC/GIS and
XIS, we estimated the XIS flux for both Center/ER and WR observations
with {\tt xissimarfgen} by using the same normalized PSPC image and
assumed spectrum, confirming that the XIS flux and the estimated one
are consistent within 10\%.  

Second, we estimated the flux and spectral shape of the CXB from the
observational results of \citet{kirsch05}.  Specifically, we model the
CXB spectrum by a powerlaw with a photon index of 1.4 with a cut-off at 40
keV. The surface brightness is again folded through the HXD field of
view, yielding 6.9$\times$10$^{-12}$ erg s$^{-1}$ cm$^{-2}$ in 15$-$50
keV.

Third, we used the  ROSAT 2RXP catalogue
\footnote{\tt ftp://ftp.xray.mpe.mpg.de/rosat/catalogues/2rxp/pub}
order to model the contribution of point sources.  This catalogue
allows us to search for point sources over the whole HXD field of view.
We found 2 (4) sources in the HXD-PIN field of the Center/ER
(WR) observation.
% although it covers only lower energy (0.2$-$2.35 keV)
Assuming a powerlaw spectrum with photon index of 1.5 and the 2RXP
normalization, the total point source contribution to each pointing is
quite small, 5.8 (8.2) $\times$10$^{-13}$ erg s$^{-1}$ cm$^{-2}$
(15$-$50 keV) for the Center/ER (WR) region. The flux from
obscured AGNs not detected by ROSAT is likely to also be small, as
discussed in \citet{nevalainen04}.  Since the current PIN response
gives a factor of 1.13 larger model normalization than the XIS
response (\cite{kokubun07}), we multiplied the flux of all the above
components by this factor.

\begin{figure}[ht]
\begin{center}
\begin{minipage}[tbhn]{8.0cm}
\FigureFile(7.5cm,7.5cm){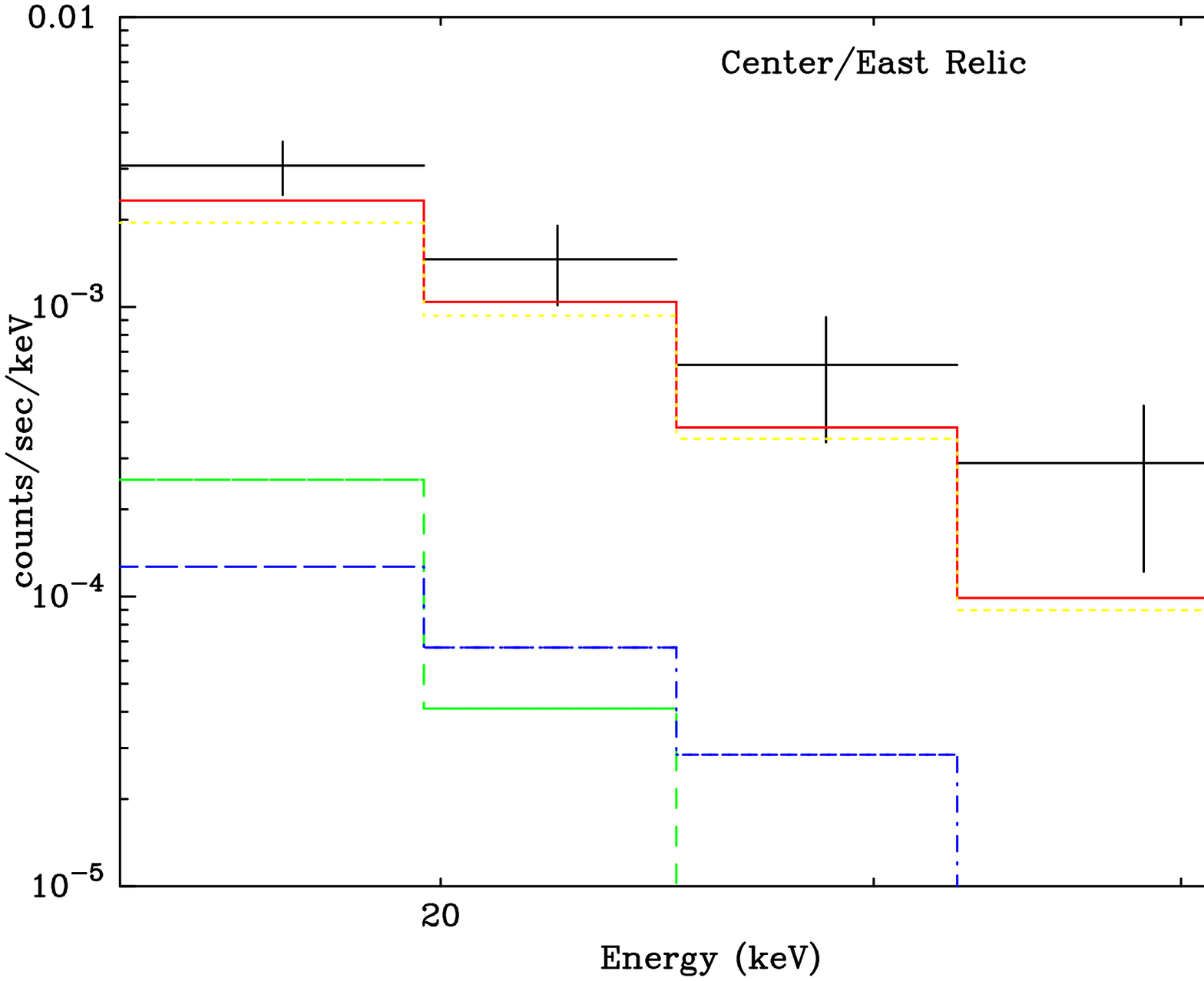}
\end{minipage}\quad
\begin{minipage}[tbhn]{8.0cm}
\FigureFile(7.5cm,7.5cm){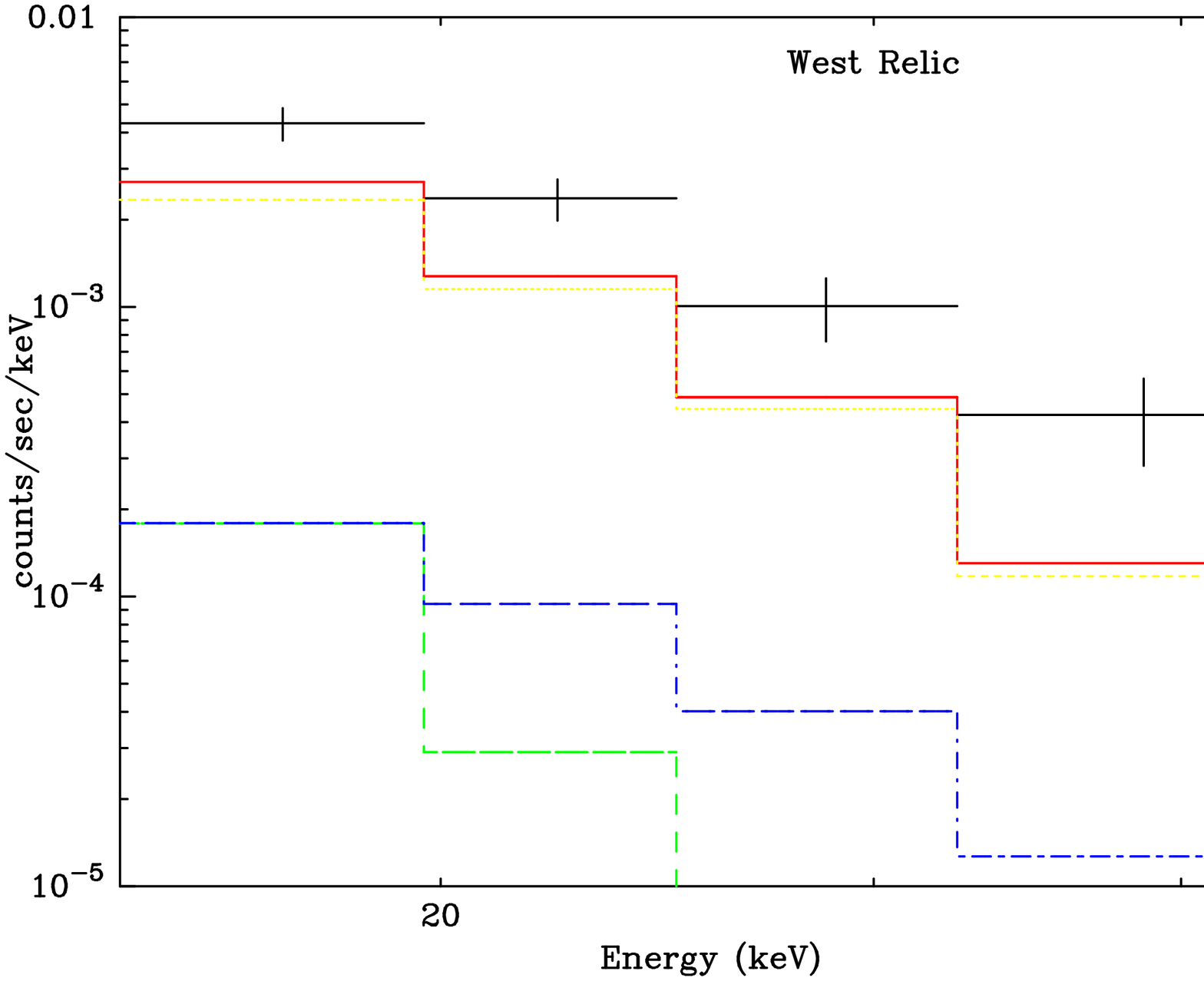}
\end{minipage}
\end{center}
\caption{NXB-subtracted HXD PIN spectra from Abell 3376 Center/ER (left) and 
WR (right). The black crosses show NXB-subtracted data while the solid lines 
correspond to the contribution of CXB (yellow), thermal ICM (green), point 
sources (blue), and their sum (red), respectively.}
\label{pinspec}
\end{figure}

Figure \ref{pinspec} shows the NXB-subtracted spectra of the HXD PIN,
together with each model component.  We find a weak hard X-ray excess
above 15 keV for Abell 3376 WR.  When we assumed the excess emission
to be a power-law with a photon index of 2.0, its flux is (7.2 $\pm$
2.4) $\times$10$^{-12}$ erg s$^{-1}$ cm$^{-2}$ (15$-$50 keV) at the 90\%
confidence level after the
flux of the other components are excluded.  

We estimated the systematic uncertainty of each spectral component in
order to determine whether the excess is significant. Fluxes are
reported in the 15$-$50 keV range hereafter. The PIN NXB has a
systematic uncertainty of 2 \% at the 1$\sigma$ level as mentioned
above. This gives the largest uncertainty, which corresponds to
$\sim$2.8$\times$10$^{-12}$ erg s$^{-1}$ cm$^{-2}$.  For the ICM
emission, the uncertainty of the flux (i.e. $normalization$) comes
from modeling the surface brightness profile and assumption of the
uniform temperature distribution.  However, the ICM emission is much
softer and fainter than the excess hard X-ray emission, and thus the
ICM flux uncertainty cannot explain the excess emission.  The
systematic uncertainty of the CXB ($\sigma_{\rm CXB}$) arises from two
reasons; an intrinsic intensity fluctuation on the sky
\citep{kushino02} and a systematic uncertainty of the normalization in
past measurements.  The fluctuation is calculated by scaling the
HEAO-1 result of 2.8 \% with the equation $\sigma_{\rm CXB} \propto
\Omega^{-0.5} S^{0.25}$\citep{condon74}, where $\Omega$ = 15.8 deg$^2$ and $S \sim 8\times10^{-11} \ \rm erg cm^{-2} s^{-1}$ are the effective
solid angle of the observation and upper cut-off flux of the
detectable discrete sources in the field of view, respectively. In the
case of the HXD ($\Omega = 34' \times 34'$, and $S \sim 8\times
10^{-12} \ \rm erg cm^{-2} s^{-1}$), the systematic fluctuation of the CXB is
calculated to be 9.2\% (1$\sigma$) of CXB flux.  
The normalization uncertainty is
around 10\% when we survey past measurements \citep{kirsch05},
corresponding to $\sim$0.8$\times$10$^{-12}$ erg s$^{-1}$ cm$^{-2}$.
Finally, for point sources, we consider the sum of the statistical
errors of each point source obtained from ROSAT 2RXP catalogue as the
systematic uncertainty. This gives a fluctuation of at most $\sim$0.04
$\times$10$^{-12}$ erg s$^{-1}$ cm$^{-2}$ (1$\sigma$).  
In summary, the dominant error source is the uncertainty of the 
PIN NXB estimation.
We combined these systematic errors by summing squares.

Considering all the systematic uncertainties, the hard X-ray flux of
A3376 WR is (7.2 $\pm$ 4.8 $\pm$ 8.8) $\times$10$^{-12}$ erg s$^{-1}$
cm$^{-2}$ at the 3$\sigma$ confidence level, 
where a photon index of 2.0 is assumed.  Therefore, the
non-thermal X-ray emission is not significant.  The 3$\sigma$ upper
limit is 2.1$\times$10$^{-11}$ erg s$^{-1}$ cm$^{-2}$, by adding both
the statistical and systematic errors at that level.  In the same
manner, the hard X-ray flux of
A3376 ER is (3.2 $\pm$ 2.4 $\pm$ 8.8) $\times$10$^{-12}$ erg s$^{-1}$
cm$^{-2}$ at the 3$\sigma$ confidence level, 
and the 3$\sigma$ upper-limit flux for the Center/ER observation
is derived to be 1.4$\times$10$^{-11}$ erg s$^{-1}$ cm$^{-2}$; .  The
flux detected with BeppoSAX/PDS is (8.0$\pm$8.9)$\times10^{-12}$
(3$\sigma$ error) when scaled to the energy band of 15--50 keV by a
power-law model with a photon index of 2.0 (\cite{nevalainen04}), and
thus the upper limit is similar to our results.  Note that our
constraint per unit solid angle, $1.8\times10^{-14}$ erg cm$^{-2}$
s$^{-1}$ arcmin$^{-2}$, is for a smaller sky region of $34' \times
34'$ around the WR, while the BeppoSAX upper limit per unit solid
angle, $2.8\times10^{-15}$ erg cm$^{-2}$ s$^{-1}$ arcmin$^{-2}$, is
for a larger sky region of $78' \times 78'$.
%When the officially archived NXB is utilized, it gives a similar
%situation; only the upper limit around 
%$<$1.2$\times$10$^{-11}$ erg s$^{-1}$ cm$^{-2}$ could be obtained. 

\subsubsection{XIS Spectra\label{s-xissp}}

The highest energy channels of the XIS are useful for searching for
non-thermal emission in the periphery of clusters of galaxies since
the XIS achieves the lowest NXB among the existing CCD detectors. We
here present the results from the XIS spectral analysis.  We first
created the XIS NXB spectra by means of the background generator {\tt
xisntebgdgen}, in which the NXB spectrum for the same detector region
as the on-source spectra is estimated from the night earth observation
database, based on the history of the COR in each observation.  The
reproducibility of the XIS NXB seems to be somewhat different among
the four sensors (XIS0$-$3), and it is reported to be 2.8$-$4.4 \% (1
$\sigma$) in the 5--12 keV band \citep{tawa07}.

The CXB spectrum is modeled by a power-law with a photon index of 1.4.
We utilized the spectrum of the west region of WR, where the cluster
emission is negligible, to measure the CXB normalization.  We
generated the auxiliary response files with {\tt xissimarfgen},
assuming that the incident emission is spatially uniform and comes
from a circle with 20 arcmin radius, combined with the integration
region on the detector for the above spectrum.  The spectrum at the
west region of WR is completely consistent with the model in the high
energy band, but additional low-temperature thermal emission, which
is due to the Galactic emission, is needed in the lower energy band.
We fitted these components by adding a two-temperature APEC plasma
model of solar abundance, obtaining temperatures of 0.08 keV and 0.28
keV and the normalization for both.  Then, we modeled the emission for
the on-source region of the A3376 integration by preparing the auxiliary
response file for that region and folding the above CXB and Galactic
emission model through the XIS. We subtracted the thus-obtained
background spectra and the NXB spectrum from the on-source spectra.

\begin{figure}[ht]
\begin{center}
\begin{minipage}[tbhn]{8.0cm}
%\rotatebox{0}{\FigureFile(8.0cm,6.0cm){fig3a.ps}}
\rotatebox{0}{\FigureFile(8.0cm,6.0cm){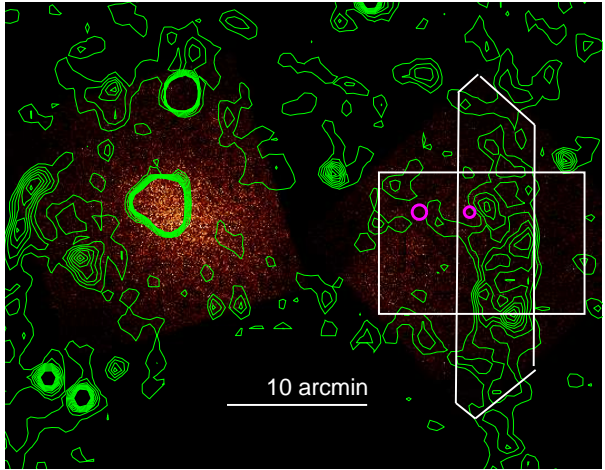}}
\end{minipage}
\end{center}
\caption{Suzaku XIS image of Abell 3376 in the 4--8 keV band. All the images of
 XIS0$-$3 are merged. The 1.4 GHz radio contours of NVSS are superposed. Two
 magenta circles indicate the positions of point sources in the XIS field
 of view. The square represents the region of projection analysis. The
 hexagon represents the integration region for spectral analysis of the
 West Relic.}
\label{image-radio}
\end{figure}

\begin{figure}[ht]
\begin{center}
\begin{minipage}[tbhn]{8.0cm}
\rotatebox{0}{\FigureFile(9.5cm,9.5cm){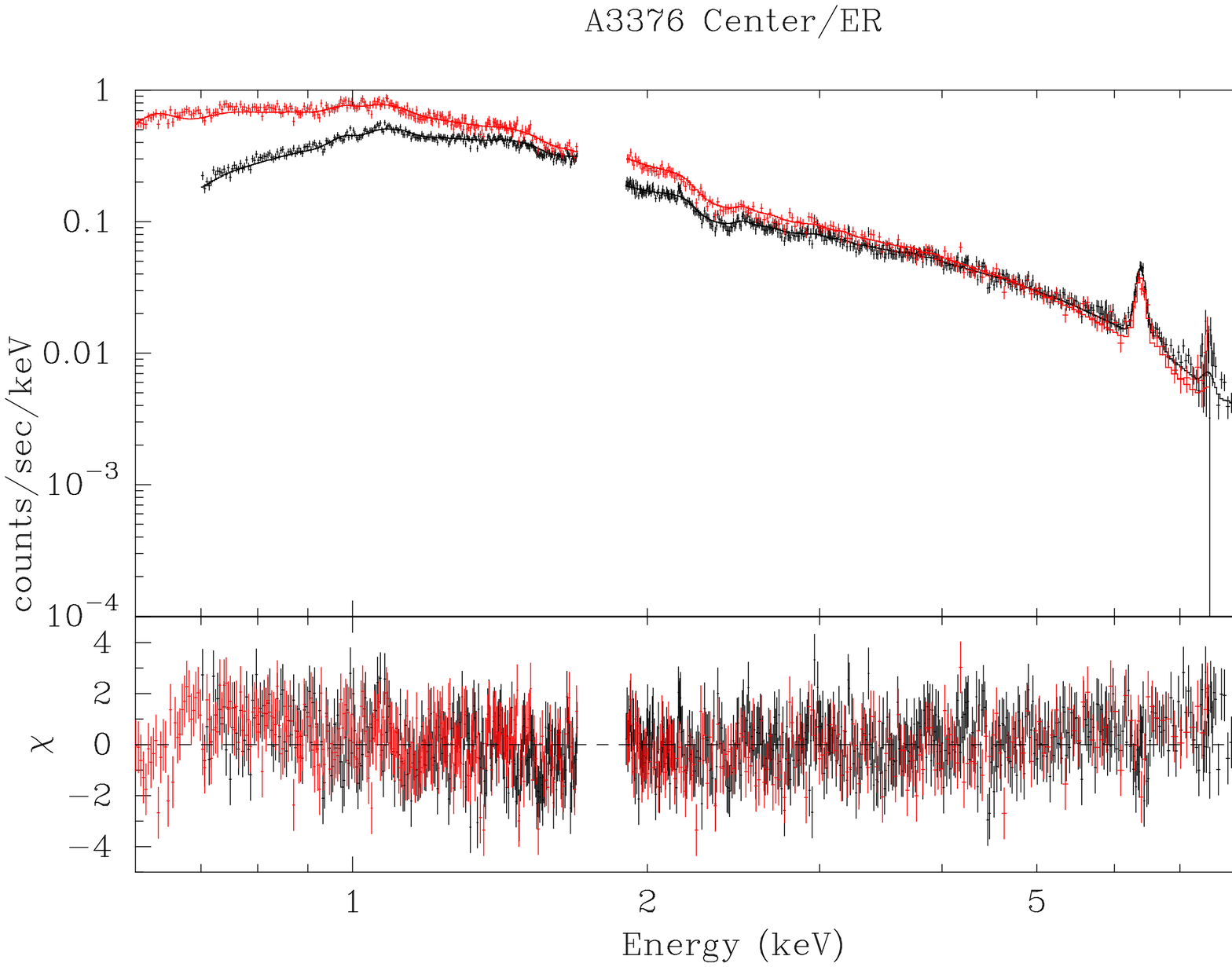}}
\end{minipage} \quad
\begin{minipage}[tbhn]{8.0cm}
\rotatebox{0}{\FigureFile(9.5cm,9.5cm){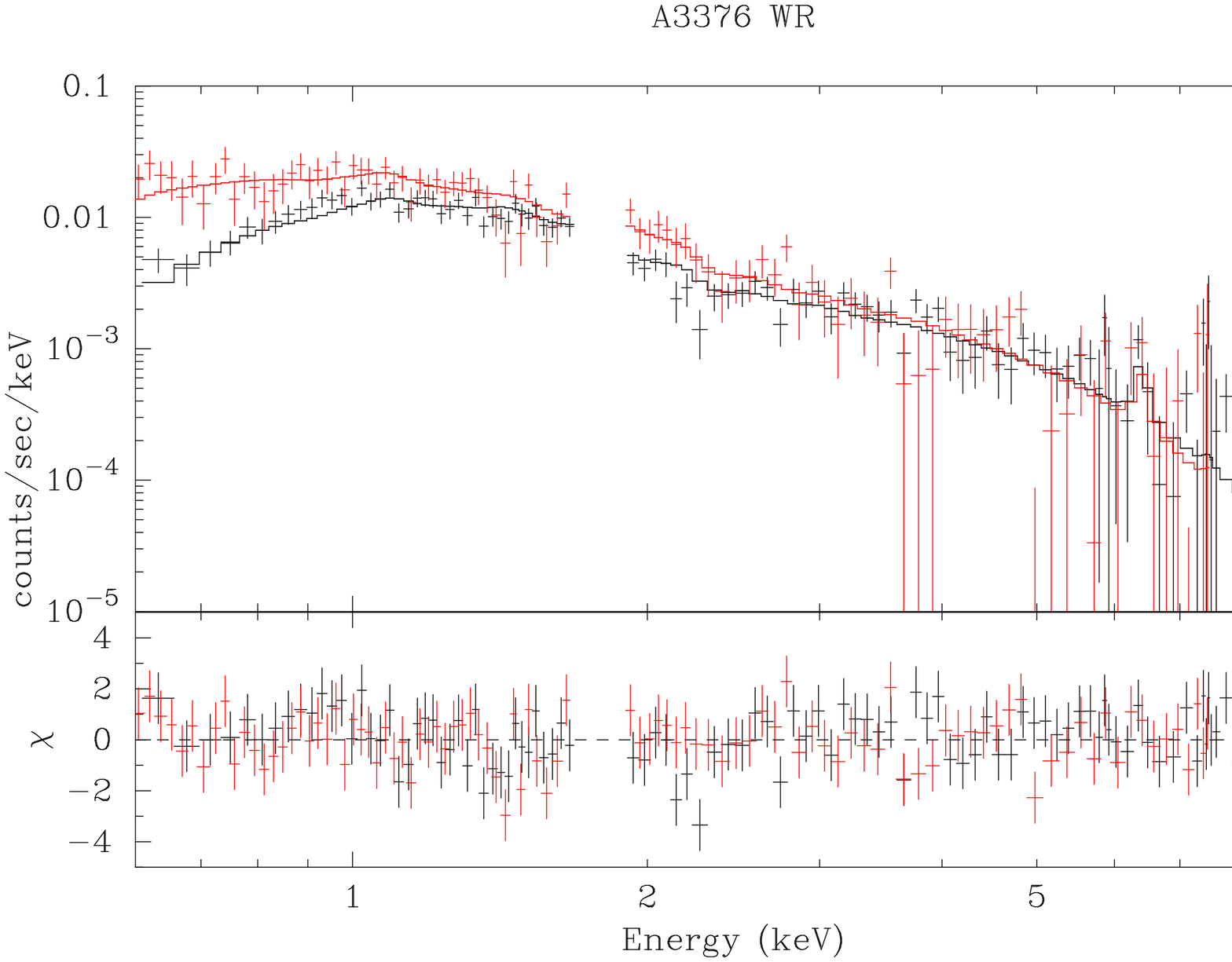}}
\end{minipage}
\end{center}
\caption{Spectral fitting of XIS spectra from Abell 3376 Center/ER (left)
 and WR (right), with the thermal plasma model. 
The red and black represent the XIS-BI and FI, respectively.
The crosses show background-subtracted data. The solid
 lines represent the best-fit model.}
\label{xisspec}
\end{figure}

Figure \ref{image-radio} shows the hard band image in the 4$-$8 keV
band for both Abell 3376 Center/ER and WR. A blank-sky image,
extracted from the Lockman Hole observed on May 7, 2006, is subtracted
as the background data from Abell 3376 source image.  The 1.4 GHz
radio contours are superposed. The cluster in the hard band exhibits
an elongated appearance similar to that in the total band (Figure
\ref{image}) or low energy band image, and there is no remarkable
feature associated with the radio emission.  The spectra of Abell 3376
are extracted from the full field of view of the XIS after excluding
the calibration source region for the Center/ER observation, and from
a hexagon region with a width of 6$'$ covering the radio relic for
the WR observation (Figure \ref{image-radio}).  Using the ROSAT PSPC
image, we estimated the stray light from the bright cluster center to
the WR region with {\tt xissimarfgen}, and found that it is only 25\%.
%We ignored this effect in the following analysis,
%since stray contribution itself is already included in the arf file
%and the center and the WR spectra are similar.
The stray light contribution itself is already included in the arf
files and the center and the WR spectra are similar.  Before
constraining the nonthermal emission, we first fitted the Abell 3376
spectra with a single temperature APEC plasma model, without an extra
power-law component.  The summary of the spectral fitting is shown in
table \ref{xis-fitpara} and figure \ref{xisspec}.  The ICM temperature
and metal abundance of the Center/ER region are roughly consistent
with the ASCA and XMM-Newton results (\cite{fukazawa04},
\cite{bagchi05}).  On the other hand, the temperature in the WR region
is somewhat lower.  Next we included a power-law model to represent
the nonthermal emission.  The photon index is assumed to be 2.0 in the
same manner as in the PIN analysis.  The power-law component is
significantly required for the Center/ER spectra, considering only the
statistical error.

We consider the 3$\sigma$ systematic error arising from the
uncertainties of the response matrix, background model, and ICM
emission model, since the CXB fluctuations are at most 6\% (3$\sigma$) of the
total flux.  The first is 0.6 (0.05)$\times$10$^{-12}$ erg s$^{-1}$
cm$^{-2}$ for the Center/ER (WR), by assigning a 10\% error to the
response matrix \citep{serlemitsos07} and taking 10\% of the observed
total flux.  For the second, we took a systematic uncertainty of the
XIS NXB as 10 \% at the 3$\sigma$ level, fitted the spectra by
renormalizing the background by 0.9 and 1.1, and obtained the error
from the allowed power-law flux range; 0.2 $\times$10$^{-12}$ erg
s$^{-1}$ cm$^{-2}$ for both the center/ER and WR observation.  The
residual around 7 keV in the Center/ER spectra almost disappears when
the NXB level is raised by 5\%.  We estimate the error on the third by
adding another APEC component to the model, since this cluster is
merging and possibly consists of a multi-temperature structure.  For
the center/ER spectra, this operation significantly improves the fit
and the residual in the higher energy band almost disappears; the
systematic error is 2.0 (0.29) $\times$10$^{-12}$ erg s$^{-1}$
cm$^{-2}$ for the Center/ER (WR).

Taking into account the above systematic uncertainties, the flux of
the power-law component is constrained to be (1.1 $\pm$ 0.4 $\pm$ 2.0)
$\times$10$^{-12}$ erg s$^{-1}$ cm$^{-2}$ and (4.6 $\pm$ 4.0 $\pm$
2.9) $\times$10$^{-13}$ erg s$^{-1}$ cm$^{-2}$ (4--8 keV) for the
center/ER and WR, respectively, at 3$\sigma$ level.  The detection of
both observations is not significant, and therefore we obtain a
3$\sigma$ upper limit on the power-law flux in the 4--8 keV band:
$<$3.5$\times$10$^{-12}$ erg s$^{-1}$ cm$^{-2}$ (Center/ER), and
$<$1.1$\times$10$^{-12}$ erg s$^{-1}$ cm$^{-2}$ (WR).  When they are
extrapolated to the PIN energy band, the upper limit in the 15--50 keV
band is $<$6.1$\times$10$^{-12}$ erg s$^{-1}$ cm$^{-2}$ (Center/ER),
and $<$1.9$\times$10$^{-12}$ erg s$^{-1}$ cm$^{-2}$ (WR).  This
constraint gives a surface brightness per unit solid angle of
$<1.9\times10^{-14}$ erg s$^{-1}$ cm$^{-2}$ arcmin$^{-2}$ (Center/ER),
and $<1.5\times10^{-14}$ erg s$^{-1}$ cm$^{-2}$ arcmin$^{-2}$ (WR),
similar to that of the upper limit of HXD-PIN for the WR,
$1.6\times10^{-14}$ erg s$^{-1}$ cm$^{-2}$, when considering the
difference of the field of view.

Since the Fe-K line may be affected by the non-thermal component, we
fitted the XIS spectrum only around the Fe-K band (6--7 keV) with a
single-temperature plasma model.  The WR spectrum is too poor to
perform such an analysis, and thus we here analyzed only the center/ER
spectrum.  We see a weak signature of the H-like Fe-K$\alpha$ line.
However, the obtained temperature of $4.6\pm0.5$ keV is consistent
with the value of the whole-band spectral fitting.  No significant
deviation of the line ratio between He-like and H-like lines is
observed.  The upper limit of the Fe-K line broadening is 1200 km
s$^{-1}$.

\begin{table}[ht]
\caption{Best-fit parameters for the XIS spectral fitting}
\label{xis-fitpara}
\begin{center}
%\small
\begin{tabular}{cccccc}
\hline\hline
Position & Temperature$^{\ast}$ & Abundance$^{\ast}$ & Reduced $\chi^2$
 $^{\ast}$ & Reduced $\chi^2$ $^{\ddag}$ &  $F_{\rm po}^{\dag}$ \\
\hline
Area & (keV) & (solar) & ($\chi^2$/d.o.f.) & ($\chi^2$/d.o.f.) 
 & (erg s$^{-1}$ cm$^{-2}$) \\
\hline
Center/East relic    \\
$18\times18$ arcmin$^2$        & 4.12$\pm0.05$ 
                               & 0.29$\pm0.01$ 
                               & 1.32 (1581/1201) 
                               & 1.07 (1281/1200) 
                               & $(1.1\pm0.2)\times$10$^{-12}$ \\
West Relic           \\
$122$ arcmin$^2$               & 3.81$\pm0.33$ 
                               & 0.22$\pm0.10$ 
                               & 1.08 (171/158) 
                               & 1.00 (157/157) 
                               & $(4.6\pm2.0)\times$10$^{-13}$ \\
\hline
\multicolumn{6}{l}{$\ast$  : Results for the model that
 does not contain the nonthermal power-law component.} \\
\multicolumn{6}{l}{$\ddag$  : Reduced $\chi^2$ for the model that
 contains the nonthermal power-law component.} \\
\multicolumn{6}{l}{$\dag$  : Allowed flux of the power-law
 component in 4$-$8 keV, when it is included in the spectral
 fitting. }\\
\multicolumn{6}{l}{\qquad Here only the statistical error is shown.} \\
\end{tabular}
\end{center}
\normalsize
\end{table}

\subsection{Comparison of X-ray and Radio Image around the West Relic}

\begin{figure}[ht]
\begin{center}
\begin{minipage}[tbhn]{8.0cm}
\rotatebox{-90}{\FigureFile(7.0cm,8.0cm){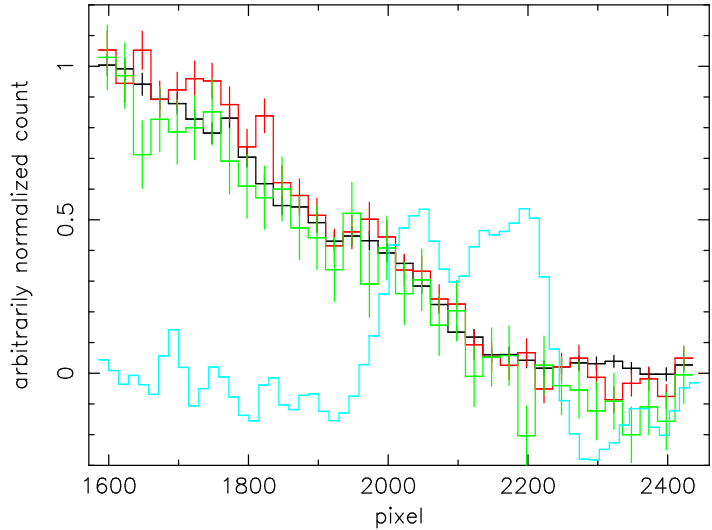}}
\end{minipage}
\end{center}
\caption{X-ray count rate profiles in the 1$-$2 (black), 2$-$4 (red), and
 4$-$8 (green) keV bands along the east-west direction, extracted from the
 white square region in figure \ref{image-radio}. The 1.4 GHz radio
 profile is also plotted (light blue). Two point
 sources are located at pixels 1750 and 1970.}
\label{image-ewdist}
\end{figure}

In order to compare the X-ray and radio image quantitatively, we
extracted the count rate profiles along the east-west direction from
the white square region in figure \ref{image-radio}, and show the
profiles in figure \ref{image-ewdist} along with the 1.4 GHz radio
profile. The X-ray profile is divided into three energy bands; 1$-$2,
2$-$4, and 4$-$8 keV. The count rate is arbitrarily scaled so as to
distinguish the difference among profiles easily. Two magenta circles
in figure \ref{image-radio} show the positions of two point sources in
the field of view. These are at pixels 1750 and 1970 in the
profiles, and correspond to local enhancements in the projected count
rate. The X-ray emission exhibits similar profiles among the three
energy bands.  The count rate ratio between 4--8 keV and 1--2 keV band
is $0.94\pm0.06$ and $0.83\pm0.21$ in the range 1700--1900 pixels
(off-WR) and 2000--2200 pixels (on-WR), respectively.  This is
consistent with the insignificant detection of nonthermal emission in the
spectral analysis.  In addition, we note that there are no X-ray
decrement in any energy band at the WR, indicating that the nonthermal
pressure of thr WR is much weaker than that of thermal pressure.

\section{Discussion}

Suzaku has observed Abell 3376 twice and constrained the non-thermal
hard X-ray emission.  An upper limit on the excess emission above the
thermal component obtained from the HXD PIN spectrum from a 34$'
\times$34$'$ (FWHM) region is $<$1.4$\times$10$^{-11}$ erg s$^{-1}$
cm$^{-2}$ and $<$2.1$\times$10$^{-11}$ erg s$^{-1}$ cm$^{-2}$ (15$-$50
keV) for the center/ER and the WR, respectively.  These are similar to
the BeppoSAX upper limit (\cite{nevalainen04}).  The XIS hard band
spectrum of a 122 arcmin$^2$ region around the WR also gives a tight
upper limit of $<$1.1$\times$10$^{-12}$ erg s$^{-1}$ cm$^{-2}$ (4$-$8
keV).  Thanks to the narrower field of view of the XIS and PIN
compared to the BeppoSAX/PDS, some constraints on the energy density
of the magnetic field and relativistic electrons can be obtained for a
more limited region around the radio West Relic.

\begin{figure}[ht]
\begin{center}
\rotatebox{0}{\FigureFile(9.5cm,9.5cm){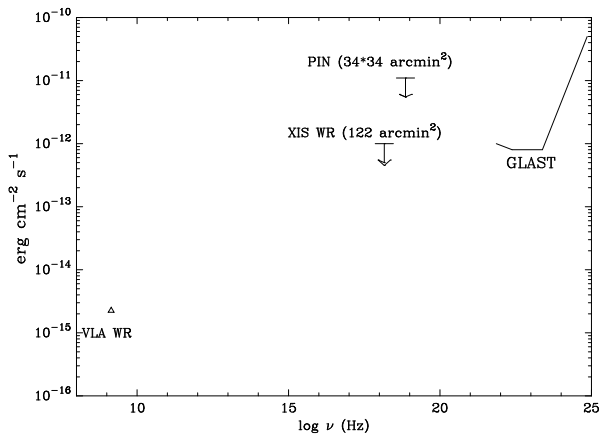}\quad}
\end{center}
\caption{Plot of the radio synchrotron and X-ray upper flux of A3376
 WR non-thermal emission on the Spectral Energy Distribution. The
 sensitivity limit of GLAST is also shown.}
\label{sed}
\end{figure}

Assuming that the non-thermal emission originates from inverse Compton
scattering of the CMB photons by relativistic electrons with energy of
$\gamma\sim10^{4-5}$, where $\gamma$ is the Lorentz factor, we can
estimate a lower limit on the magnetic field by combining the X-ray
results here with the radio results.  The 1.4 GHz integrated flux
density of the Abell 3376 WR is 82$\pm$5 mJy ((\cite{bagchi02}.  Then
the radio synchrotron luminosity becomes $L_{\rm sync} =
1.9\times10^{40}$ erg s$^{-1}$, when integrated over a frequency range
of 1$-$5 GHz, assuming a photon index of $\Gamma = 2.0$.  The energy
density of the CMB is $U_{\rm CMB} = 4.2\times10^{-13} (1+z)^4 =
5.0\times10^{-13}$ erg cm$^{-3}$.  Meanwhile, the upper limits of the
IC X-ray emission are $L_{\rm HXR}<2.4\times10^{44}$ erg s$^{-1}$ for
the HXD (15$-$50 keV), and $L_{\rm HXR}<2.2\times10^{43}$ erg s$^{-1}$
for the XIS (2$-$10 keV), under the assumption that the photon index
is 2.  These are shown in figure \ref{sed}, together with the GLAST
sensitivity limit.  Using the relation $L_{\rm HXR}/L_{\rm sync} =
U_{\rm CMB}/U_{\rm B}$ (where $U_{\rm B} = \frac{1}{8\pi}B^2$), the
magnetic field is estimated to be $B >$0.03 $\mu$G for the HXD and $B
>$0.10 $\mu$G for the XIS, which do not rule out the equipartition
condition of $B\sim$ 0.5--3 $\mu$G \citep{bagchi06}.  The XIS gave a
significant constraint on non-thermal X-ray emission at the WR and
yields an upper limit on the energy density of relativistic electrons
in the WR of $9.0\times10^{-13}$ erg cm$^{-3}$ or $2.9\times10^{-13}$
erg cm$^{-3}$, assuming the electron energy index of 3 or 2,
respectively, and the minimum electron energy of $\gamma\sim10^2$,
below which the Coulomb losses are very fast.  The thermal pressure
around the WR is estimated from the temperature (3.8 keV) obtained by
the XIS in this region and the thermal electron density
($\sim1\times10^{-4}$ cm$^{-3}$) obtained with ASCA/GIS
\citep{fukazawa04}, giving $\sim1\times10^{-12}$ erg cm$^{-3}$.  Since
there is no apparent X-ray decrement at the WR, the non-thermal
pressure is likely smaller than the thermal pressure.  In this case,
the steep energy index of relativistic electrons around 3 is not
preferred if the minimum electron energy extends to $\gamma=10^2$.
The upper limit from the HXD gives an independent measurement of the
total amount of relativistic electrons not necessarily emitting at 1.4
GHz radio band, but possibly at longer wavelengths.  In the same way
as the XIS case, the non-thermal pressure by the relativistic
electrons averaged over a cube of 2.7 Mpc on a side (34' at the
distance of A3376) is less than $4.6\times10^{-13}$ erg cm$^{-3}$ or
$6.2\times10^{-14}$ erg cm$^{-3}$, assuming tan electron energy index
of 3 or 2, respectively.  These are much smaller than the thermal
pressure.  This constraint is also valuable information if the
magnetic field is strong only in the WR but the relativistic electrons
are widely distribute around the WR.

Other mechanisms of non-thermal X-ray emission that we can consider
are non-thermal bremsstrahlung and synchrotron.  Since the former is
very inefficient and will cause a large heat input into the ICM rather
than emitting hard X-rays (e.g. Petrosian 2001), we think it will not
be the dominant process for producing non-thermal hard X-rays in this
particular cluster.  X-rays from the synchrotron process need very
high energy electrons with $\gamma\sim10^8$.  Since their life time is
very short, around $10^5$ year, primary electrons accelerated in situ
are not realistic.  Secondary electrons decaying from primary protons
are possible candidates.  The current upper limit on the X-ray to
radio flux ratio allows an electron energy index of $<$2.  In this
case, GeV gamma-rays from decays of $\pi^0$s produced through protons
and intracluster hot gas is expected and GLAST can detect such
emission \citep{blasi02}, without violating the HXD result.

In the next decade, hard X-ray observations with much higher
sensitivity above 10 keV will become available from implementing new
technologies such as the hard X-ray imager planned for the NeXT
mission.  
%Imaging can lower the detectable flux by means of higher
%sensitivity and an additional hard X-ray imaging capability, compared
%to the Suzaku/HXD.  
Therefore, detection of hard X-ray emission is promising unless the
emission is very extended.

\section{Conclusion}

Suzaku has observed Abell 3376 to search for non-thermal emission in
clusters of galaxies. Thanks to the low background of the HXD PIN, a
tight upper limit on the possible non-thermal flux is obtained for the
Abell 3376 WR region. The XIS produced an upper limit on
the non-thermal flux in the soft energy band (4$-$8 keV) for the first
time.  Although future improvements in the NXB reproducibility will yield
increased sensitivity, our result already demonstrates the potential
of the XIS and HXD/PIN to search for non-thermal emission from galaxy
clusters.

Authors thank to Dr. J. Nevalainen for careful reading and many useful 
comments.
NK is supported by Japan Society for the Promotion of Science 
Postdoctoral Fellowship for Young Scientists.
JPH acknowledges support from NASA grant NNG06GC04G.

%%%
% See the manual for the detail.
%%%

\appendix
\section{PIN NXB Estimation}

We give a full description of the PIN NXB estimation used in this paper in
\citet{kawano06}. Here we briefly summarize it.

\subsection{Estimation Method\label{nxb}}

\begin{figure}[ht]
\begin{center}
\begin{minipage}[tbhn]{8.0cm}
\rotatebox{0}{\FigureFile(8.0cm,6.0cm){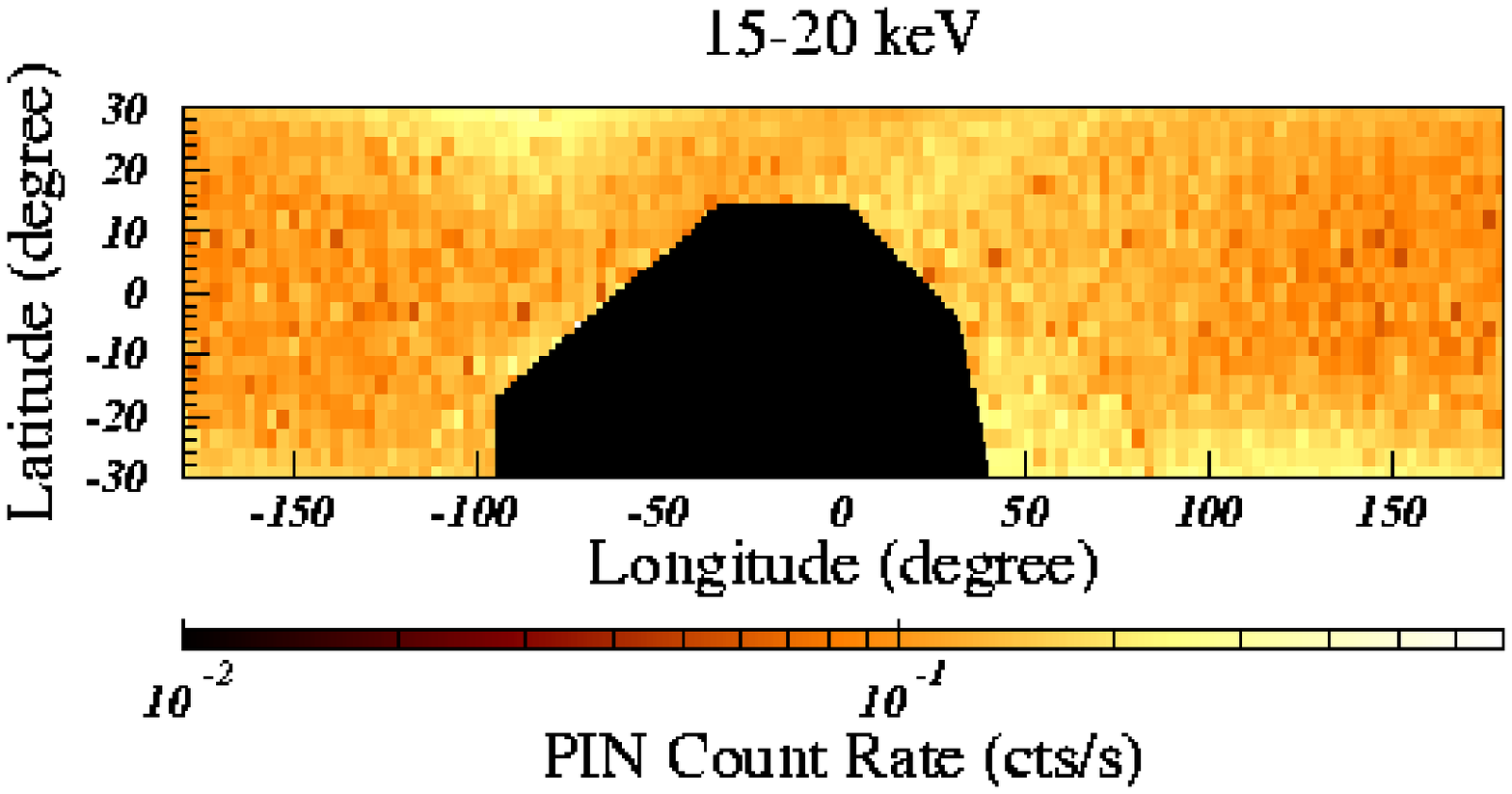}}
\end{minipage} \quad
\begin{minipage}[tbhn]{8.0cm}
\rotatebox{0}{\FigureFile(8.0cm,6.0cm){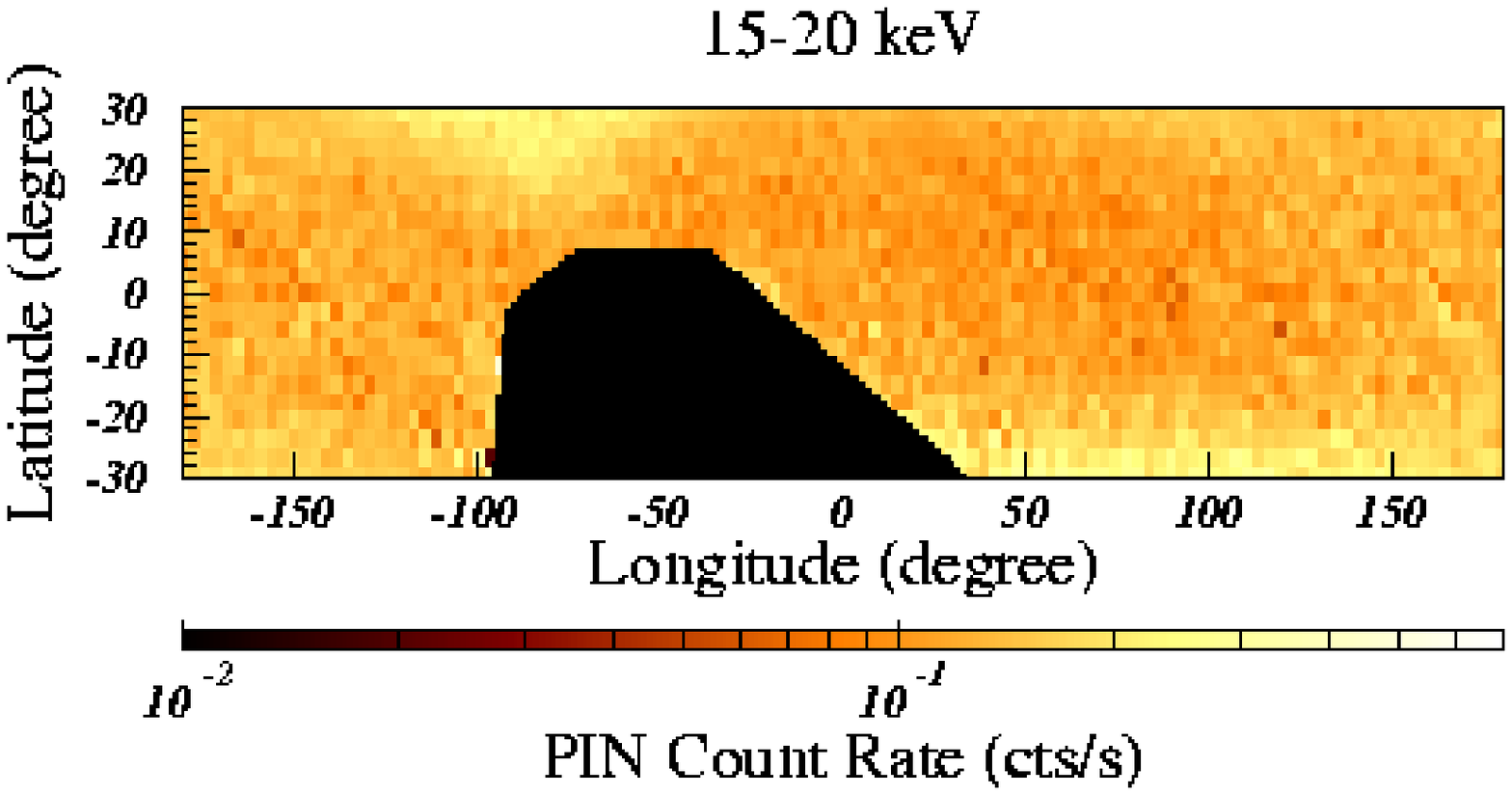}}
\end{minipage}
\end{center}
\caption{Examples of PIN NXB maps. Left and right are in the 15--20
 keV band, with and without SAA passage, respectively. One
 pixel is 3$\times$3 degree$^2$. Filled black area in each
 map corresponds to the SAA region.}
\label{pinmap}
\end{figure}

\begin{figure}[ht]
\begin{center}
\begin{minipage}[tbhn]{8.0cm}
\rotatebox{-90}{\FigureFile(6.0cm,6.0cm){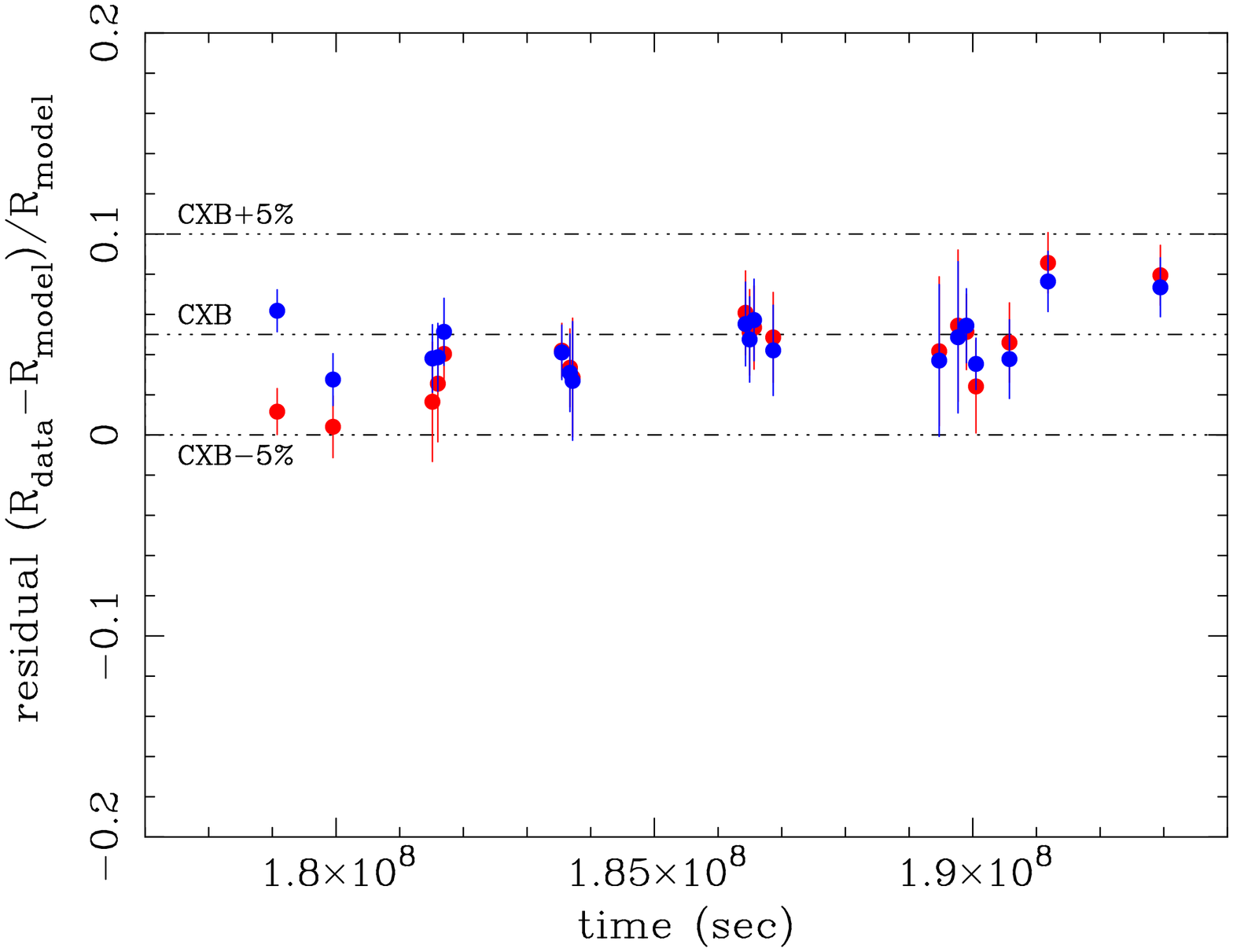}}
\end{minipage} \quad
\begin{minipage}[tbhn]{8.0cm}
\rotatebox{-90}{\FigureFile(5.5cm,6.0cm){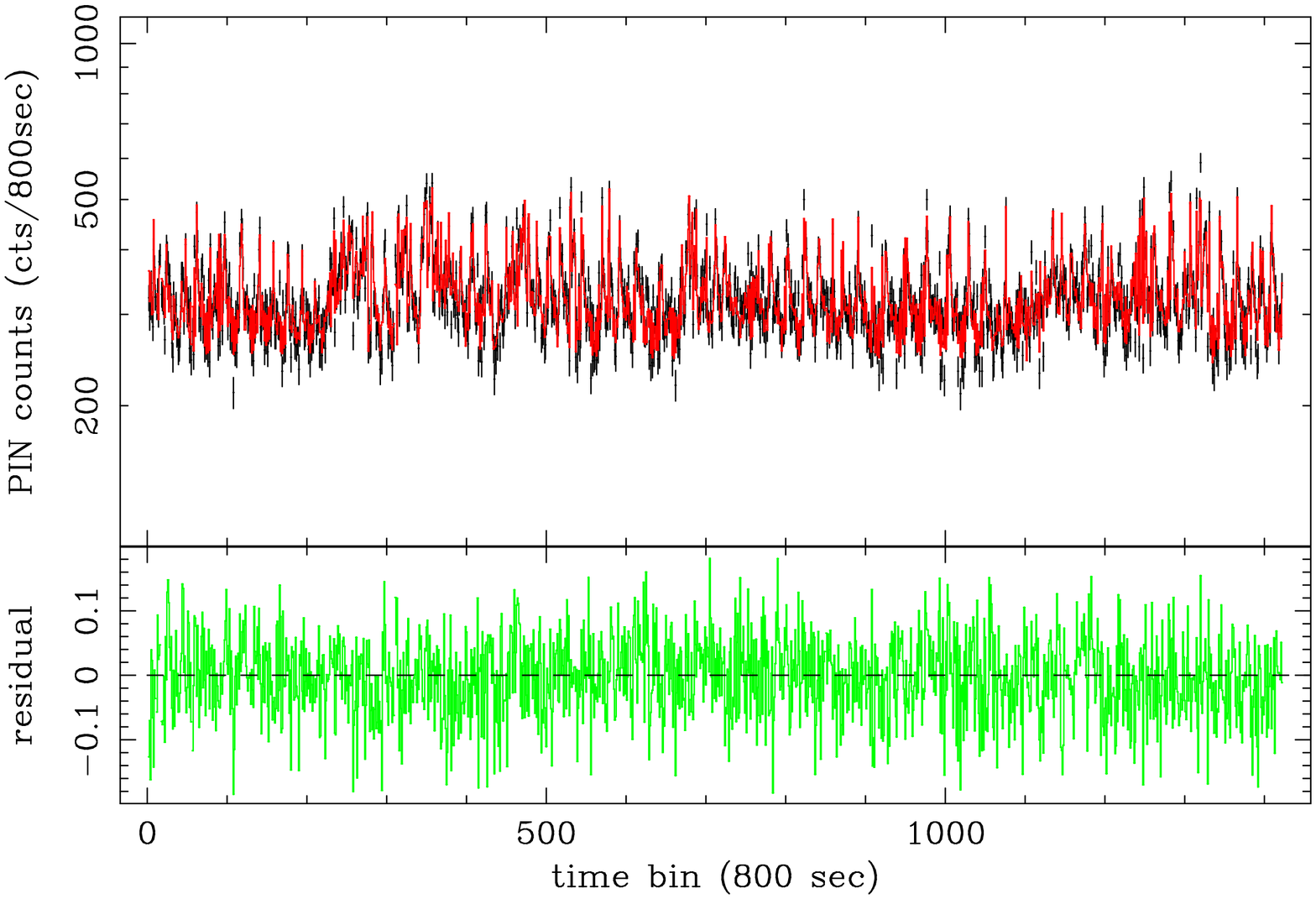}}
\end{minipage}
\end{center}
\caption{The left figure shows the time dependence of the fractional 
 residual between the dark earth data and the NXB model. 
 Red points are for the NXB
 estimation from the raw map data, and blue points are after the
 long-term variation of the NXB is corrected. Three dot-dashed lines 
 represent the
 level of 0.95, 1.00, and 1.05 of the CXB, which is itself only 0.05 of the
 NXB. The right top figure shows the light curve of the PIN NXB data (black) 
 fitted with the emperical model (red) over 6 months. 
 See the text for details. The right bottom panel
 shows the fractional residual between the NXB data and the model of the PIN 
 NXB (green). Horizontal axis represents the order of time-sorted data bins.}
\label{residual-lcfit}
\end{figure}

The NXB of the PIN detector depends strongly on the COR and the
activation after the SAA passage \citep{kokubun07}.  COR is the
shielding power against charged particles in units of GeV c$^{-1}$,
and the best parameter to predict the PIN NXB with good
accuracy. Since the COR changes as the satellite's position above the
earth changes, we first create PIN NXB maps using the earth
occultation data from September 2, 2005 to March 11, 2006.  The total
exposure time is $\sim$2 Msec.  The data are divided into 16 maps in
eight energy band (0$-$11, 11$-$15, 15$-$20, 20$-$25, 25$-$32.5,
32.5$-$45, 45$-$70, 70$-$90 keV) and two kinds of orbital path (with
or without SAA passage). Each map is binned by 3$\times$3 degree$^2$
so that one pixel typically contains 100$-$200 counts.  Examples of
maps for SAA or non-SAA passage orbits in the 15$-$20 keV band are
shown in figure \ref{pinmap}.  The clear difference of NXB pattern due
to activation in the SAA can be seen.  When the time and position
(i.e. latitude and longitude) are input, we can estimate the PIN NXB
from these maps.  Figure \ref{residual-lcfit} left (red circles) shows
the fractional residuals between the estimated PIN NXB and the
observational data on sources sufficiently faint to be undetectable
with the PIN.  The signal in these observations can be regarded as CXB
plus NXB.  It is obvious that the residuals increase by $\sim$10 \% as
time goes by.  This implies that it is necessary to take into account
the long-term variation of PIN NXB.

Based on the above, we then go back to the NXB (e.g. earth
occultation data) light curve and fit it with an empirical model 
including four time-variable component as follows,
\begin{eqnarray}
 BGD(t) &=& A \times PINUD^{2} + B \times PINUD \nonumber \\
        &+& C \times \left( \int PINUD\exp (-t_{SAA}/\tau )dt \right)^D \nonumber \\
        &+& E \times (1 + \sin (2 \pi \times 2.3\times10^{-7} \times (t - F))) \nonumber \\
        &+& G \times (1 - H \times \exp (-t/I)) \nonumber \\
        &+& J \ \ (constant). \nonumber \\
\end{eqnarray}
Here, capitals from A to J stand for the fitted variables.  The first
two terms give the COR dependence, which is well traced by the PINUD
count.  PINUD is a charged particle monitor included with the HXD.
The terms with parameters C to I express the activation component.
The short-term component is estimated by the PINUD build-up count;
convolution of the PINUD count by an exponential function with an
appropriate time constant (12500 sec).  The middle-term component is
found to produce an approximately periodic variation with $\sim$50
days, as a result of the modulation of SAA activation due to the
precession of the Suzaku orbit, and thus it is represented by the sine
curve.  The term with G, H, and I represents the long-term component.
We measure the free parameters A to J from fitting the light curve
over the entire PIN energy range of 11-90 keV. We fit separately for
those orbits with/without SAA passage. The light curve without SAA
passage is shown in figure \ref{residual-lcfit} right. The $\chi^2$
root-mean-square (RMS) residuals with (without) SAA passage are
1774/1009 = 1.76 (1878/1413 = 1.33), and $\sim$6.8 (6.3) \%,
respectively. The contribution of each component for the total NXB
counts is (a) $\sim$73 \% for the COR-dependent terms, (b) $\sim$22 \%
for short-term, (c) $\sim$5 \% for middle-term, and (d) $<$1 \% for
long-term activations, respectively.  We correct the template PIN NXB
maps by the effects of only components (c) and (d), since components
(a) and (d) are already incorporated into the map data bases.  After
this procedure the long-term trend of the residuals in figure
\ref{residual-lcfit} left is reduced by this procedure.  We generate
an 8 energy-band NXB model spectrum for any observation using the
corrected PIN NXB map and the actual satellite trajectory of the
observation on the map.

\begin{figure}[ht]
\begin{center}\begin{minipage}[tbhn]{8.0cm}
\rotatebox{-90}{\FigureFile(8.0cm,6.0cm){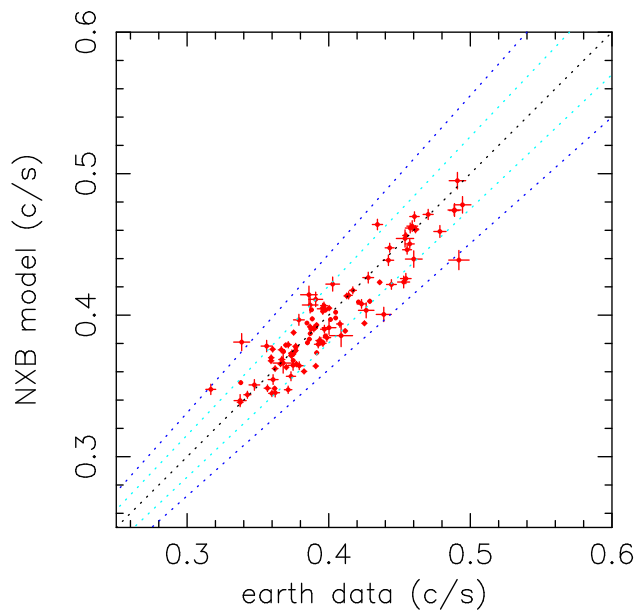}}
\end{minipage}
\end{center}
\caption{Comparison of count rate between the estimated NXB and the real 
dark earth data for 100 observations. Dotted lines represent 5 \% (light blue)
and 10 \% (blue) deviation level.}
\label{cts-dist}
\end{figure}

In order to confirm the reproducibility of the estimated PIN NXB, we
compared the count rates predicted by the NXB model with real dark
earth data for 100 observations each of whose exposure is longer than
3000 sec (i.e. more than $\sim$1000 cts). We furthermore chose an exposure of
$>30000$ sec since data of shorter exposure suffer a large statistical
error. As shown in Figure \ref{cts-dist}, the model and data are in
good agreement within $\sim$5 \%.  We made the distribution histogram
of $\sum_i\frac{\sqrt{\left(d_i^2-m_i^2\right)-e_i^2}}{m_i}$, where
$d_i$, $m_i$, $e_i$ are the real NXB count rate, model NXB rate, and
real NXB statistical error.  We fitted the histogram with a Gaussian
whose Gaussian parameter $\sigma$ is 2\%.  In other words, a
systematic error in our NXB estimation is $\sim$2\% at 1$\sigma$.

\subsection{Check of the Estimated PIN NXB for the Abell 3376 Center/ER 
Observations\label{nxb2}}

\begin{figure}[ht]
\begin{center}
\begin{minipage}[tbhn]{8.0cm}
\rotatebox{-90}{\FigureFile(6.0cm,6.0cm){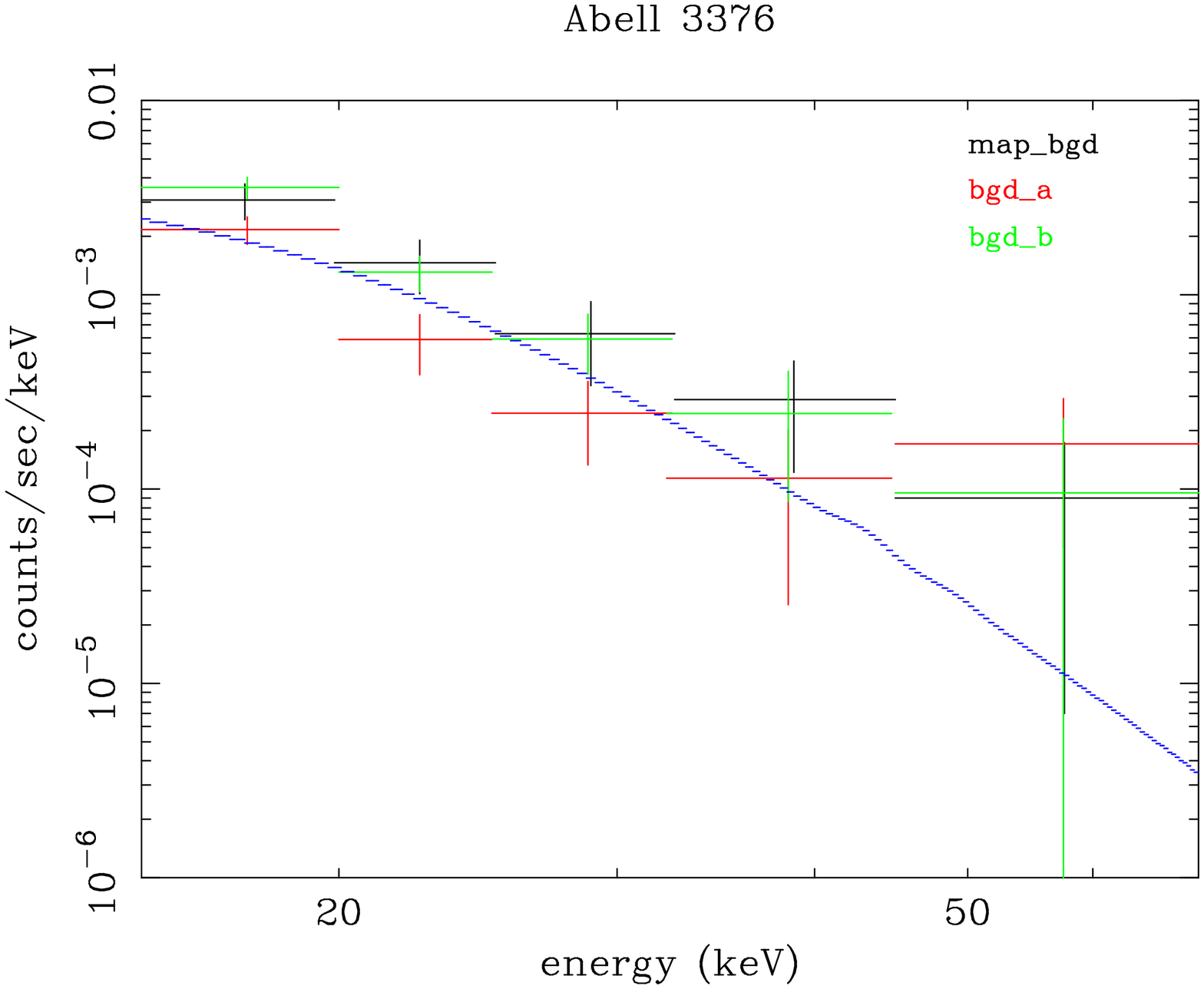}}
\end{minipage} \quad
\begin{minipage}[tbhn]{8.0cm}
\rotatebox{-90}{\FigureFile(6.0cm,6.0cm){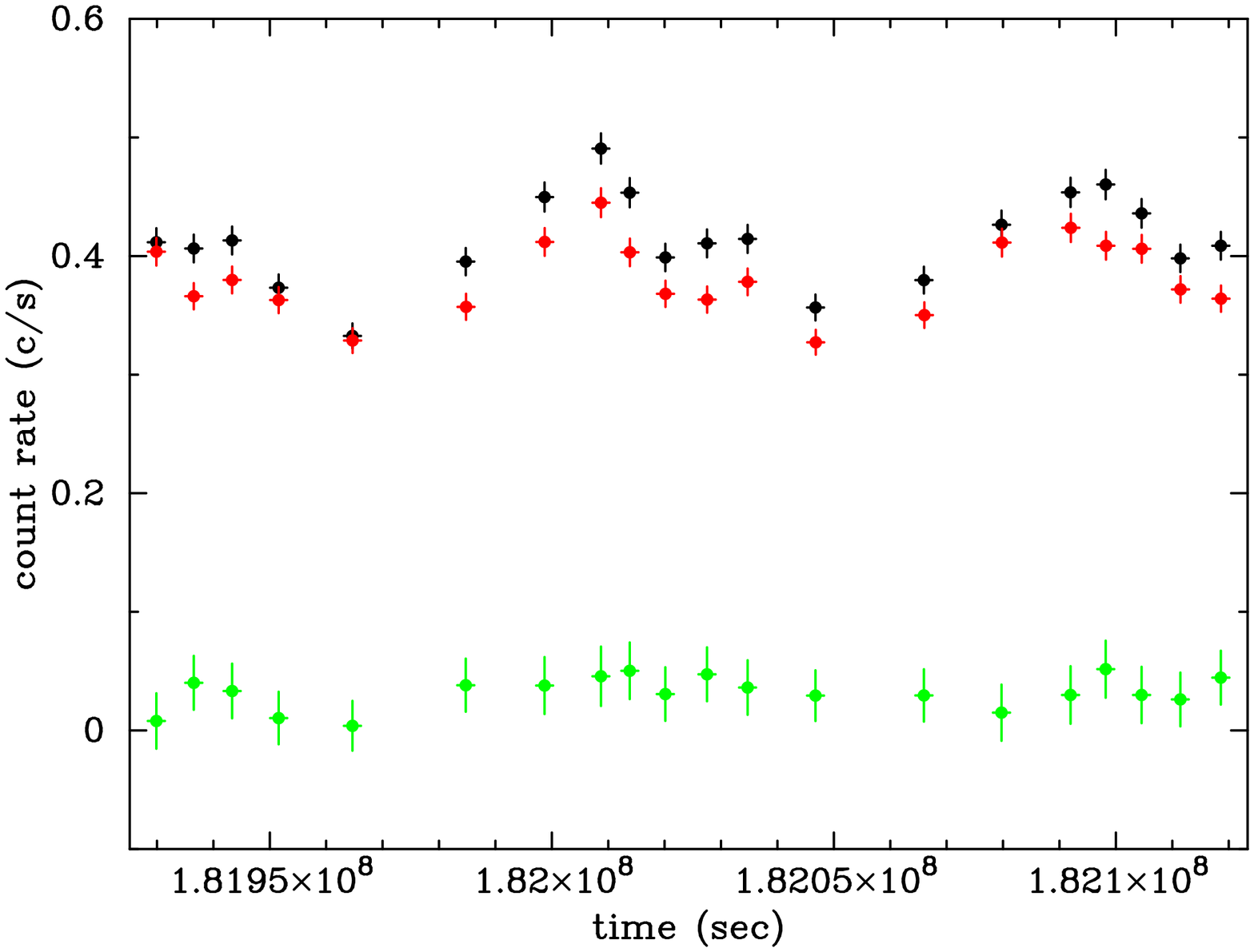}}
\end{minipage}
\end{center}
\caption{Comparison of background-subtracted spectra (left) and light
curve (right) of the A3376 center/ER region (Oct. 6--10, 2005). Black
points in the left panel are the spectra from which the NXB estimated
by our method is subtracted. Red and green points use the officially
supplied NXB models of bgd\_a and bgd\_d, respectively.  the CXB level
is given by the blue line. In the right panel black, red, and green
data points are the total count rate, the NXB model with our method,
and the background-subtracted count rate, respectively.}
\label{cmp-bgda}
\end{figure}

We here check the estimated PIN NXB in two different ways.  Apart from
our PIN NXB model, there are two other models (bgd\_a, bgd\_d)
developed and supplied by the HXD team.  The first (bgd\_a) mainly
uses the PINUD rate and the PINUD build-up with a fixed decaying time
constant. The second (bgd\_d) estimates the NXB with an emperical
time-variation model, which is obtained by fitting the light curve of
the earth occultation data with a more complex model than that in the
previous subsection.  As described in \S3.1, the NXB estimation of the
above two models is not accurate for the observation of the WR region.
On the other hand, they are available for the observation of the
center/ER region.  Therefore, we can check the consistency between
three model NXBs. Figure \ref{cmp-bgda} left shows the comparison of
the background-subtracted spectra.  They are almost consistent within
statistical errors.  The spectrum using bgd\_a gives a somewhat low
count rate.  Such a tendency is seen in the report of background
reproducibility \citep{mizuno06} in the early phase of Suzaku
observations.

The constancy of the NXB-subtracted light curve is also useful to
check the background reproducibility for stable source such as galaxy
clusters.  The 3000 sec binned light curve of the Abell 3376 Center/ER
observation in 15--70 keV is shown in figure \ref{cmp-bgda} right.  It
is found that the light curve is constant with time within statistical
errors.  The same results are also obtained for the WR pointing and
other observations.  
%Thus, it is confirmed that the NXB reproducibility is stably good.  
We also investigated the NXB-subtracted spectra obtained in three
different periods in each observation of A3376, and confirmed
consistency among the three spectra within statistical errors.

\end{document}